\documentclass[a4paper]{article}

\usepackage[english]{babel}
\usepackage[utf8x]{inputenc}
\usepackage[T1]{fontenc}
\usepackage{lmodern,textcomp}

\usepackage[a4paper,top=3cm,bottom=2cm,left=3cm,right=3cm,marginparwidth=1.75cm]{geometry}

\usepackage{amsmath}
\usepackage{graphicx}
\usepackage{booktabs}
\usepackage{rotating}
\usepackage{lineno} 
\usepackage[colorinlistoftodos]{todonotes}
\usepackage[colorlinks=true, allcolors=black]{hyperref}
\usepackage{natbib}
\usepackage{placeins} 
\usepackage[T1]{fontenc}
\usepackage{babel}

\usepackage{caption}
\usepackage{subcaption}

\providecommand{\keywords}[1]
{
  \small
  \textbf{\textit{Keywords---}} #1
}

\usepackage{setspace}

\title{With great power (prices) comes great tail pipe emissions? \\ A natural experiment of electricity prices and electric car adoption.}
\large
\author{ Johannes Mauritzen\\
Norwegian University of Science and Technology and\\
NHH Norwegian School of Economics\\
7491 Trondheim, Norway\\
johannes.mauritzen@ntnu.no\\
}

\begin{document}

\maketitle


\abstract{\textbf{Abstract.} A fundemantal and unanswered question for the widely shared goal of electrifying passenger vehicles is how the price of electricity, which can vary greatly across countries and regions, affects buying behavior. I make use of a natural experiment in Norway in the period 2021-2022 when large price differences between north and south emerged to estimate the effect of electricity prices on the decision to purchase a pure battery-electric vehicle. Simple difference estimates along the border of the price zones as well as a difference-in-difference regression model  suggest a significant but economically modest effect of a 2-4\% reduction in the probability of purchasing an electric vehicle in the high price zone. A counterfactual simulation suggests that there would have been about 3000 to 6000 fewer electric vehicles sold in the high-price south compared to a scenario where the south had equally low prices as in the north.}

\doublespacing

\section{Introduction}
A central component of most plans for eliminating net carbon emissions is to electrify transport. Falling battery costs and large investments in electric vehicle development by both new and established automotive firms has led to a rising trend of electric vehicle sales in many major markets. 

Thanks to generous subsidies, Norway has been a leader in encouraging the purchase of new electric vehicles with no tailpipe emissions and is the country with the highest penetration of electric vehicles among new car sales in the world. Norway can be seen as a labratory for the electrification of transport, providing evidence on how the market for electric cars will develop. 

An important question in the economics and environmental science literature is how changes in consumer-facing fuel prices affect automotive choice and driving behavior at both extensive and intensive margin \citep{klier_price_2010,busse_are_2013, allcott_are_2019,bento_distributional_2009,sallee_consumers_2016,davis_estimating_2011,allcott_gasoline_2014,li_gasoline_2014,li_how_2009,donna_measuring_2021,klier_price_2010,mabit_vehicle_2014}. The bulk of the evidence tends to be that large changes in gasoline prices can significantly affect buying behavior. \citet{klier_price_2010} finds for example that the quadrupling of gasoline prices in the US between 2002 to 2007 could explain about $40\%$ of the fall in market share of the traditional US automanufacturers, who tended to produce more fuel-inefficient vehicles and light trucks compared to their foreign rivals. \citet{li_how_2009} estimates that a $10\%$ increase in gasoline prices leads to an approximate $2\%$ long-run increase in fleet fuel economy. \citet{sallee_consumers_2016} also finds evidence for gasoline prices affecting vehicle choice in the used car market.

Electric motors are inherently more efficient and electricity, per standardized energy unit, has historically been cheaper than gasoline and diesel in most developed countries, thus the energy component of an electric car's total cost of ownership will tend to be substantially lower. \citet{desreveaux_techno-economic_2020} find that the total cost of ownership in France over a 5 year period was lower for an electric car (compared to diesel and gasoline cars of a similar segment) already in 2020, despite substantially higher upfront costs. This result is mainly driven by lower fuel costs. Assuming a vehicle is driven 9500 km, a diesel vehicle has estimated yearly fuel costs of approximately 750 euro, while the equivalent costs for an EV was 204 euro.

On the other hand, deregulated electricity markets tend to have high variance with occasional large spikes in prices. The above calculations were made with an assumption of electricity prices being .13 euro per kilowatt-hour. Prices reached more than 1 euro per kilowatt-hour over the course of 2022. More so, electricity markets are regional, and large persistent price differences are common across markets. 

Identifying the effect of electricity prices on the choice of buying an electric car by comparing different regions or countries with different average electricity prices is difficult because of many potential conflating variables. Countries or regions often have differing incentives and support policies for electric vehicles as well as other factors such as culture, climate, and available infrastructure that affect both the decision to purchase an electric car and electricity prices. A region where the population has a stronger preference for environmental protection may be an unobersved variable that affects both electricity prices and electric car adoption.

In this article, I make use of price differences between zones within Norway to estimate the effect of electricity prices on choice of electric cars. Norway is part of the common Nordic electricity market, which is a zonal market where prices vary across and within countries. In certain periods, these zones can experience large price differences. The exact borders for the zones do not necessarily follow established administrative, cultural or geographic borders. The identifying assumption in this article is that at the margin, whether a potential buyer lives (and charges) in a zone can be seen as a form of random assignment. This can be a strong assumption, and I test varying restrictions on the data in order to improve the balance between high-price and low-price areas.

In addition, I make use of the fact that historical price differences between price zones in Norway have been relatively small. However in the period 2021-2022 the three southern price zones in Norway experienced record high prices in line with high continental European prices, which these markets are directly connected to through transmission cables. The northern two price zones in Norway, on the other hand, experienced relatively low prices. At times the price in southern Norway was over 100 times larger than the price in mid-Norway. Using a difference-in-difference approach, I can control for time invariate factors that may affect both the decision to purchase an electric car and electricity prices in the the different price zones.

From a theoretical standpoint, higher prices in the southern Norwegian price area may have both an income effect and a price effect on electric car purchasing, which an analysis of the total number of electric cars purchased would have a hard time disentangling. However, in my model setup I analyze the probability of choosing an electric car conditional on that a new vehicle purchase takes place. This will emphasize the price effect on the choice of an electric car. In other words, the unit of observation is a binary indicator variable of whether a new vehicle registration was electric or not.

Much of the traditional literature on vehicle choice makes use of relatively complicated multinomial choice models in order to take into account multiple purchasing options while controlling for attributes that are often associated with the variable of interest (car size and fuel efficiency, for example) \citep{beggs_assessing_1981, daziano_forecasting_2014, bhat_impact_2009, donna_measuring_2021}. My approach is intentially simpler, analyzing the choice between a pure battery-electric vehicle or a vehicle that purely or partly is powered by an internal combustion engine. This choice is of interest because it is directly related to policy debates around personal vehicles as well as firm decisions in investing in electric transport. For example, subsidies are often directly tied to this binary outcome of whether a car is purely electric or not and local and national government decisions on how much to invest in charging infrastructure are tied to how many households and businesses choose pure battery-electric vehicles. 

Hybrid and plug-in hybrid vehicles, though having in part an electric drive train, are registered as "0" in the dependent variable in my analysis. Not including plug-in hybrid vehicles, which in theory could run purely on electricity, in the electric vehicle designation is not obvious. However, because a plug-in hybrid owner has the ability to switch between electricity and petrol/diesel with little to no cost, then this would dilute the estimation of the effect of electricity prices on drive-train choice. 

The results from the model indicates a measurable but modest effect of high prices on the electric car purchasing decision. The pure difference-in-difference estimates indicate that high prices led to an approximately 2-4 percentage point reduction in the probability of choosing an electric car. A counterfactual simulation with a model that allows for varying time trends estimates that had the southern price area had equally low prices as the northern price area, then between 3,000-6,000 extra purchases would have been electric out of a total of 242,000 registered vehicles in the southern price areas in this period.

Perhaps the most notable implication of this article is how modest the effect of high electricity prices are on electric car purchasing decisions. Prices in the southern area were orders of magnitude higher--sometimes upto 100 times the price in the north--over an extended period of time. Yet in this period, the trend of increasing penetration of electric vehicles continued upwards, only at a somewhat reduced slope.

\subsection{Literature on electric car adoption and policy}

An emerging literature on the market for electric vehicles has developed alongside increasing penetrations of the vehicles. \citet{beggs_assessing_1981} provides a suprisingly early analysis of estimating the demand for electric cars, though as an example of how to estimate demand for non-existant goods based on survey data. Many studies seek to estimate the effects of government policy and both intended and unintended consequences of those policies. In an analysis of data from Ireland, \citet{caulfield_measuring_2022}, finds that electric car adoption tends to be stronger in areas with high population density and with higher average income, leading to questions about both regional- and income-distributive effects of subsidy policy. While my study is not primarily a policy analysis, I do take into account both regional and income effects in the analysis.

Another strain of literature seeks to analyse the interaction of policies. \citet{gillingham_carbon_2021} considers the combination of incentives for electric cars with carbon taxes, where they demonstrate a counterintuitive result of moderate carbon taxes leading t an electric car being charged by more carbon intensive electricity compared to no carbon tax (though that result reverses under higher carbon taxes.) One of the rationales for the electric car policy in Norway is a clean energy mix composed of primarily hydropower, though increasingly also wind power. Though Norway is connected through transmission cables with European countries with a broader mix of generation.

As opposed to the carrot-heavy incentive schemes currently in place, \citet{holland_electric_2021} constructs a model to evaluate the effects of a policy that bans gasoline and diesel vehicles in favor of electric vehicles. They find such a policy to be inefficient with a large dead weight loss. Norway currently has a non-binding goal that all light passenger vehicles should be zero-emissions (primarily electric, given current technology trends) by 2025, but no formal ban has been put into place.

Several studies have specifically analysed the Norwegian and Scandinavian policies for electrifying the passenger vehicle fleet and reducing transport emissions overall. \citet{yan_economic_2018} analyses the overall Norwegian vehicle registration tax, which was directly linked to CO2 emissions in 2007, finding that the tax was effective in shifting consumers to cars that were less CO2 intensive. The new vehicle registration tax is waived for electric vehicles, which is a main reason for why they are financially advantageous in Norway, as discussed further below. In a study of cross price elasticities between the prices of similar diesel, gasoline and electric cars in Norway, \citet{fridstrom_direct_2021} finds elasticities of .36 and .48 between electric cars and respectively gasoline and diesel cars. This relatively high degree of substitutability is suggestive that higher electricity prices may also influence vehicle choice. 

\subsection{Norwegian electric car policy and the Norwegian electricity system}

\begin{figure}
  \centering
    \includegraphics[width=.7\linewidth]{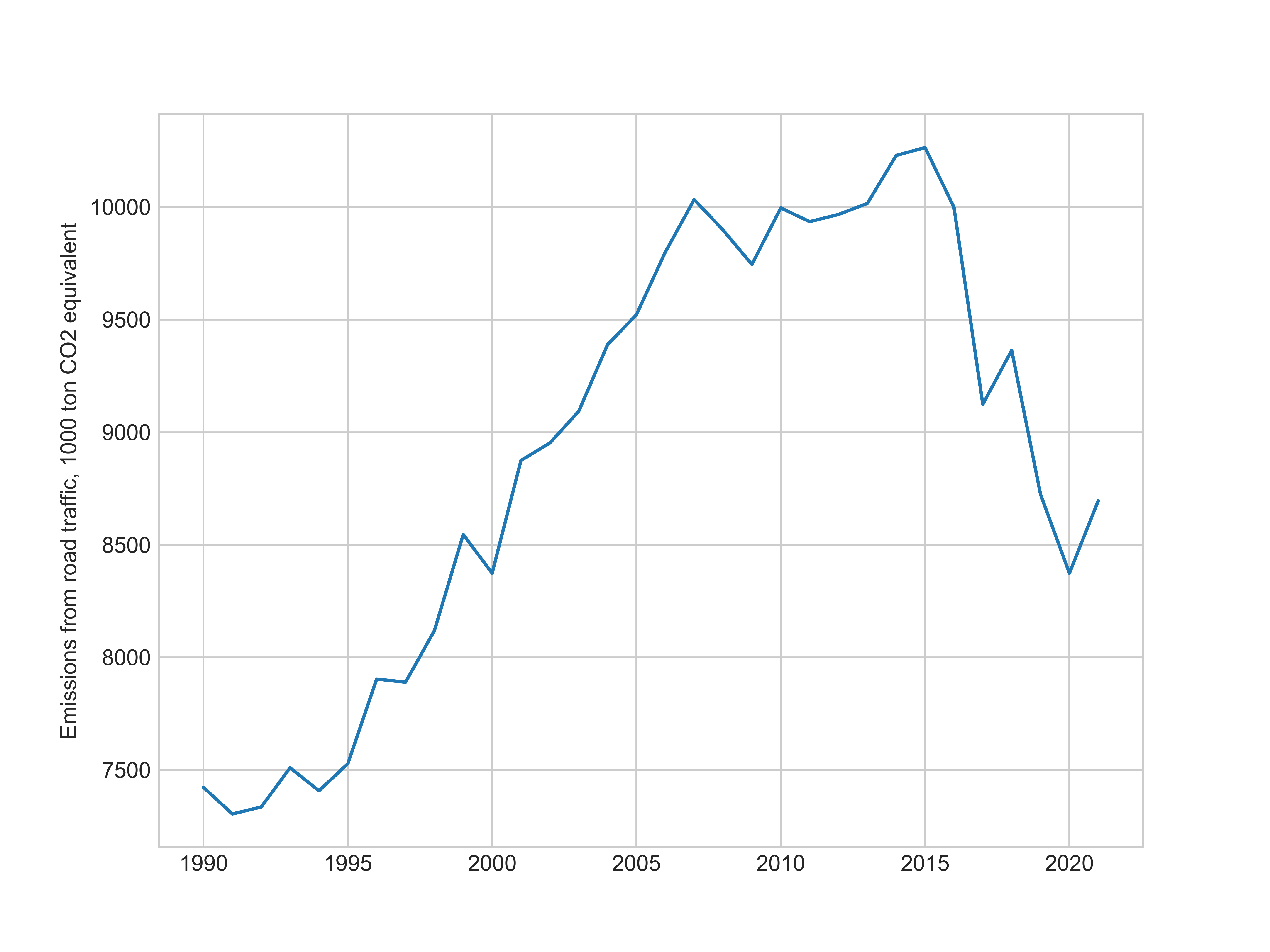}
  \caption{\label{fig:roadEmissions} Road transport emissions in Norway. Source: Statistics Norway}
\end{figure}

The Norwegian state government as well as local governments provide generous support for the purchase and operation of electric cars. The combined effect of these policies has been an early and wide-scale adoption of electric cars. By mid 2022, approximately $70\%$ of all new cars sold were electric, and approximately $18\%$ of the total passenger vehicle fleet in Norway was electric.\footnote{\url{https://www.regjeringen.no/no/tema/transport-og-kommunikasjon/veg_og_vegtrafikk/faktaartikler-vei-og-ts/norge-er-elektrisk/id2677481/}} The faster-than-expected electrification of the passenger vehicle fleet led to lower CO2 emissions from transport. Total emissions from road transport has declined more than $15\%$ from 2015 through 2021 as shown in figure \ref{fig:roadEmissions}.

At the state level, the main instruments for supporting electric cars have been a waving of the 25\% Value-Added Tax (VAT) as well as the one-time new vehicle registration fees. The registration fees can vary based on factors such as the weight of the car, engine capacity and CO2 emissions, but relative to international levels, the Norwegian fees are high. For example, a Toyota Camry with a Petrol engine\footnote{Based on the assumption that the car weighs 1500kg, the engine has a displacement of 2487 cubic centimeters and 203 horsepower with NOx emissions of 60mg/km, as entered in the tax calculator at the website of the Norwegian Tax Authority: https://www.skatteetaten.no/person/avgifter/bil/importere/regn-ut/} would have a one-time fee of more than 400,000 Norwegian kroners (Approximately 36,000 Euro). Purely based on purchase price, this has made electric cars competitive, and increasingly substantially cheaper than similar cars with an internal combustian engine. 

In addition to the implicit purchase subsidies, electric cars also receive valuable use subsidies. From 1997 through 2017 toll roads, ferry crossings, and municipal parking were free for electric vehicles. After 2018, municipalities could decide themselves whether to charge electric cars at toll roads and ferries, but the rate was capped at 50\% of the amount for an ICE car. This cap was raised to 70\% in 2023, and the Norwegian road authority has signaled that electric cars should not continue to get discounts after 2025.

\section{Data and descriptive analysis}

\subsection{data}

Data on new car registrations was obtained from the Norwegian Information Advisory for Road Traffic.\footnote{In Norwegian, "Opplysningsrådet for veitraffikken", \url{https://ofv.no}}. The data covers all new registered passenger vehicles in Nowray from January 1st 2017 through August 31st of 2022, which was the date at which the data was ordered. The original unit of observation is the number of a given vehicle model registered per day per zip-code. The dataset has a total of approximately 875,000 observations. Variables include the date of registration, the make and model of a vehicle, the engine type, the zip-code, municipality and county of the address to which the vehicle was registered.

The vast majority of the observations are single vehicle observations, but multiple vehicle observations are still common, especially for zip-codes with high population density. For the regression analysis, I want the observation unit to be a single vehicle registration. This is to simplify model choice, ease of model interpretation, and for the correct weighting of observations. Therefor, I transform the dataset to make the unit of observation a single vehicle registration by duplicating observations with multiple vehicle observations. So for example, a single observation with 3 Nissan Leafs that are registered in a given day and in a given zip code is duplicated so that the transformed data set would have 3 identical observations. Once this transformation is completed, the dataset has approximately 1.02 million observations.

Wholesale electricy price data for each price area is obtained from the ENTSO-E transparency platform.\footnote{ENTSO-E is a European umbrella organisation for transmission system operators. They provide europe-wide data freely available at the transparency website: \url{https://transparency.entsoe.eu/dashboard/show}}. I obtain data on municipality-level median income and population from Statistics Norway.\footnote{https://www.ssb.no/en} 

\begin{figure}
  \begin{minipage}{.48\textwidth}
    \centering
    \includegraphics[width=1\linewidth]{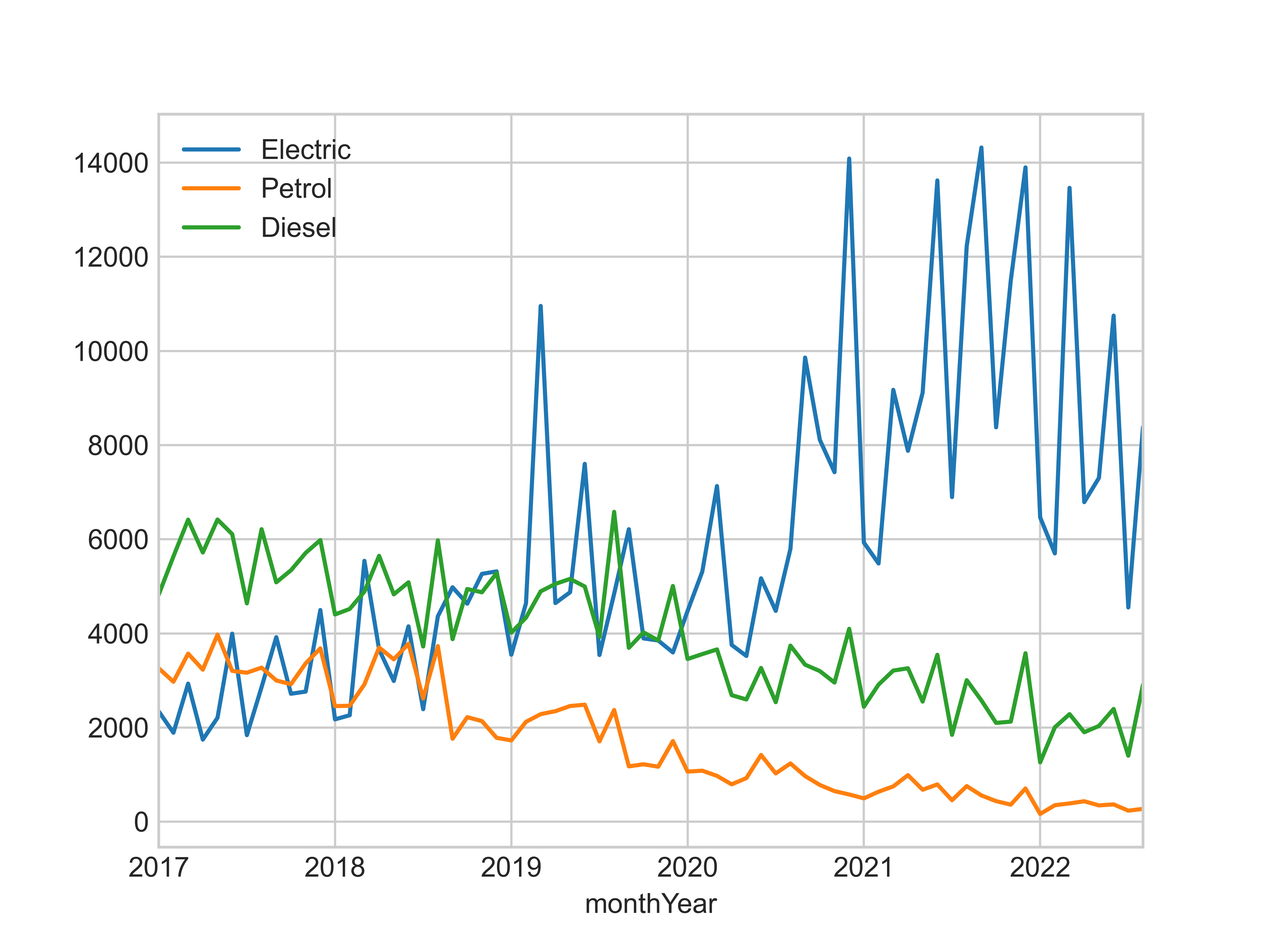}
    \caption{Cars sold by month.}
    \label{fig:monthly_carsSold}
   \end{minipage}\qquad
  \begin{minipage}{.48\textwidth}
    \centering
    \includegraphics[width=1\linewidth]{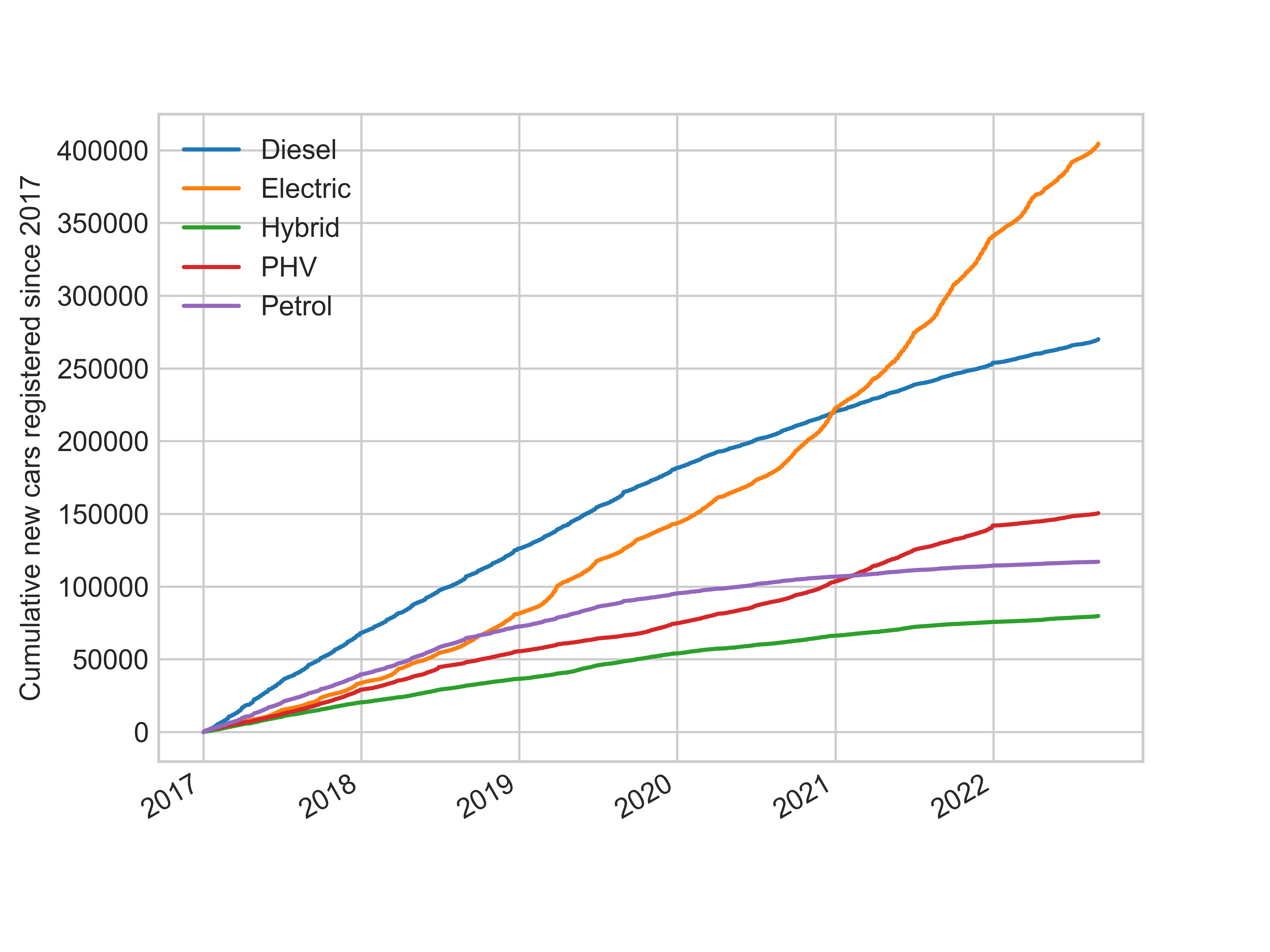}
    \caption{Cumulative cars sold.}
    \label{fig:cum_carsSold}
   \end{minipage}
  \end{figure}

Figure \ref{fig:monthly_carsSold} shows the monthly total number of electric, diesel and petrol cars sold in Norway since 2017. Already in 2017 electric cars made up a substantial portion of new cars sold, but penetration accelerated after 2020. The Covid-19 period in Norway saw a substantial increase in car purchases in part due to restrictions on use of public transport. However in this period, the number of diesel and petrol cars sold continued to decline. 

Figure \ref{fig:cum_carsSold} shows cumulative cars sold since the beginning of 2017 by drive-train type. The penetration of pure electric cars also outpaced plug-in hybrid (PHV) and hybrid cars. 

The Nordic electricity market is a zonal electricity market where uniform prices are set for pre-defined regions or in some cases whole countries. The main price mechanism is a day-ahead auction system run by the market place Nord Pool. Producers submit bid schedules for how much they are willing to produce at different price levels, and consumers--retailers and large industrial users--submit schedules of expected demand. From these bids, Nord Pool creates aggregated supply and demand schedules, which are then used to establish a theoretical system price in the market. 

\begin{figure}
  \begin{minipage}{.48\textwidth}
    \centering
    \includegraphics[width=1\linewidth]{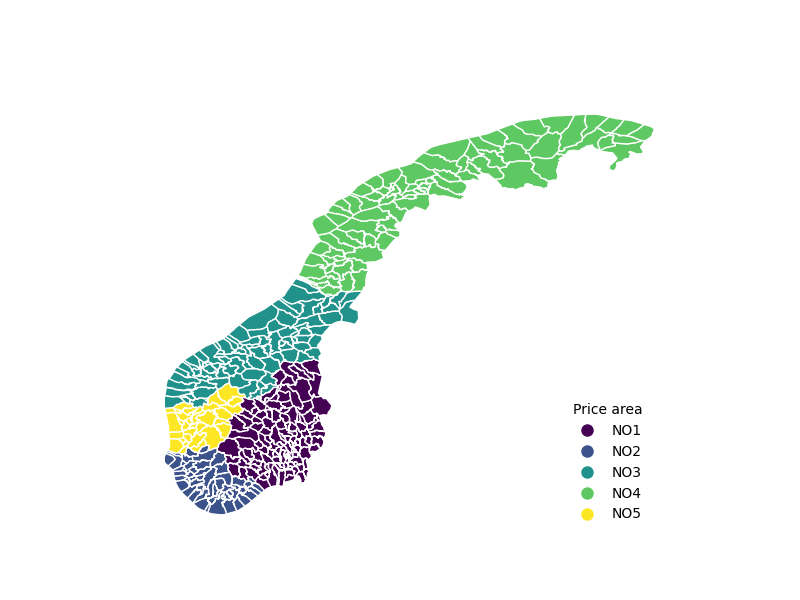}
    \caption{The map shows outlines of Norwegian municipalities coded by electricity price zones. NO3 and NO4 are the two northern-most price areas. NO1 is in the south-east, including the Oslo metropolitan area. NO2 includes southern and southwestern Norway including the cities of Kristiansand and Stavanger. NO5 includes most of the western coastal region including the city of Bergen, the second largest city of Norway.}
    \label{fig:mapPriceAreas}
   \end{minipage}\qquad
  \begin{minipage}{.48\textwidth}
    \centering
    \includegraphics[width=1\linewidth]{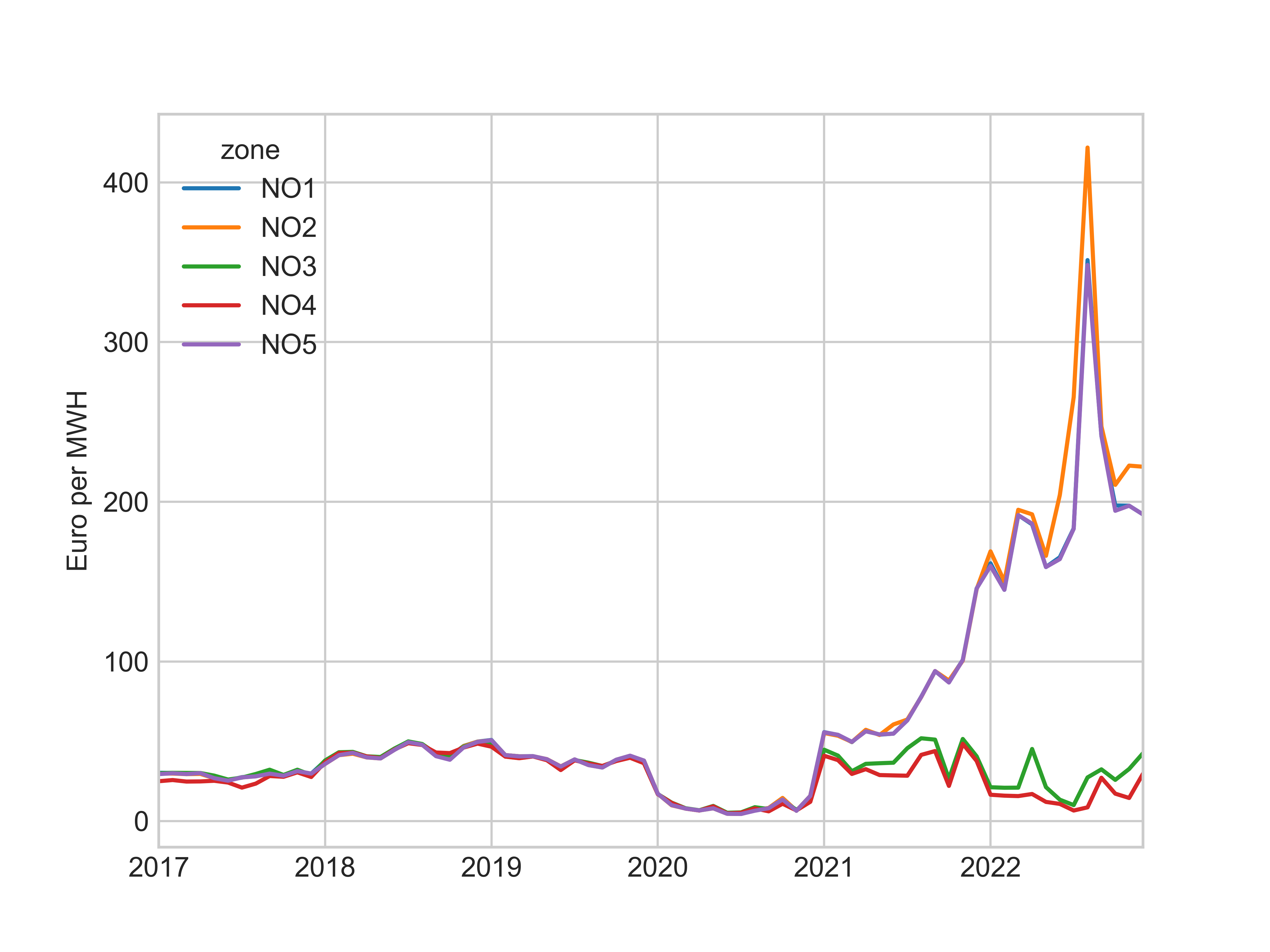}
    \caption{The figure shows monthly average prices for the five Norwegian price areas. Prices were on average similar between 2017 and 2020. Prices began to diverge in 2021, with prices in the southern price areas (NO1, NO2, and NO5) increasing substantially, while the prices in mid-Norway and northern Norway remained low.}
    \label{fig:pricesByZone}
   \end{minipage}
  \bigskip
  \end{figure}

Norway is divided into 5 price zones (often also called price areas). Figure \ref{fig:mapPriceAreas} shows a map with outlines of all municipalities with colors indicating the five price areas. NO1 covers south-eastern Norway and includes the Oslo metropolitan area. NO2 includes southern and south-western Norway and includes the cities of Kristiansand and Stavanger. NO5 covers western Norway, including Norway's second largest city of Bergen. NO3 covers mid-Norway and includes the city of Trondheim. NO4 covers the northern Norwegian area. 

Figure \ref{fig:pricesByZone} shows average monthly electricity prices in the five Norwegian price areas. Average prices were mostly similar until 2021. However, prices started diverging in the beginning of 2021, with an acceleration towards the end of 2021 and into 2022. The main driver of this divergence was the rising natural gas price in continental Europe which was often the marginal generation in continental europe and the UK. Natural gas is not used for electricity generation in Norway, but the southern price areas are all interconnected with either continental Europe and or the UK and thus the price can still be set by natural gas generation. The two northern-most price areas, on the other hand, do not have direct transmission cables to continental europe and wind and hydropower tend to be the marginal generation source. Diverging weather between south and north exasperated the trend. The eastern and southern region experienced an unusually dry and warm spring and summer in 2022, depleting hydropower reservoirs. Mid-Norway and norther Norway, on the other hand, experienced a wetter summer.

\begin{figure}
  \centering
  \caption{The map shows average yearly prices for each price area between 2020 and 2021. In 2020 there was little difference in the average price of electricity between the different price areas. Prices rose in all price areas in 2021, but more so in the southern price areas. Prices diverged in 2022 with a particularly stark contrast at the border between the mid-Norway price area (NO3) and the eastern (NO1) and western price areas (NO2).}
  \includegraphics[width=1\linewidth]{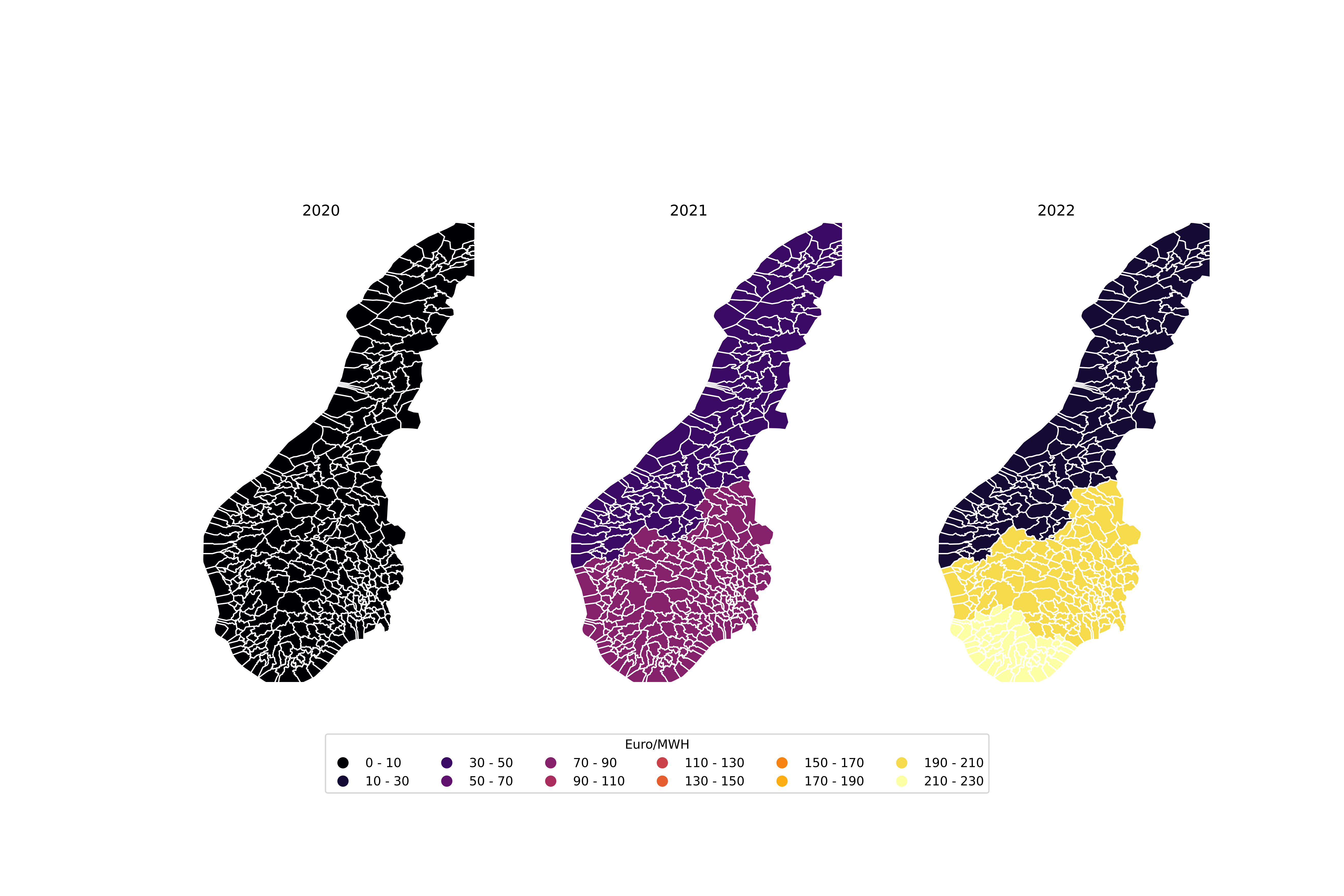}
  \label{fig:mapPrices2020_2022}
  \end{figure}

Figure \ref{fig:mapPrices2020_2022} shows average annual prices in electricity prices by price area in 2020, 2021 and 2022. The figure shows the stark discontinuity in average prices between the mid-Norway price area (NO3) and the western (NO5) and eastern Norway (NO1) prices, which is the focus of the identification strategy in this article. While the regions in Norway can be distinct at many levels--climate, economics, politics--at the margin the borders of the price areas can be seen as arbitrary. They do not necessarily follow established political or cultural borders,  cutting across counties and occasionally even municipalities.  

\begin{figure}
    \begin{subfigure}{.48\textwidth}
      \centering
      \includegraphics[width=1\linewidth]{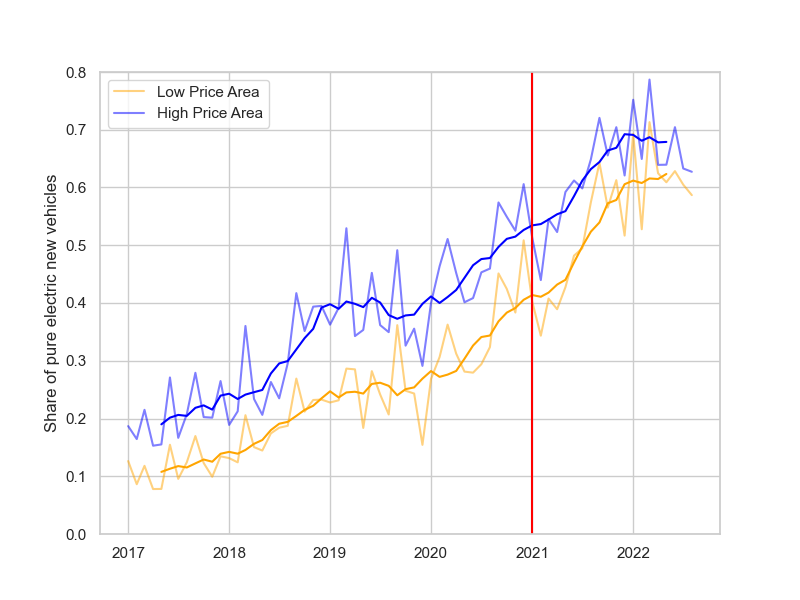}
      \caption{All Norway. Before the price divergence, the high-price area in the south had a higher penetration, though the trends appear roughly similar. After the price divergence, penetration of electric vehicles accelerates in the low-price area relative to the high price area.}
      \label{fig:shareElectric}
     \end{subfigure}\qquad
    \begin{subfigure}{.48\textwidth}
      \centering
      \includegraphics[width=1\linewidth]{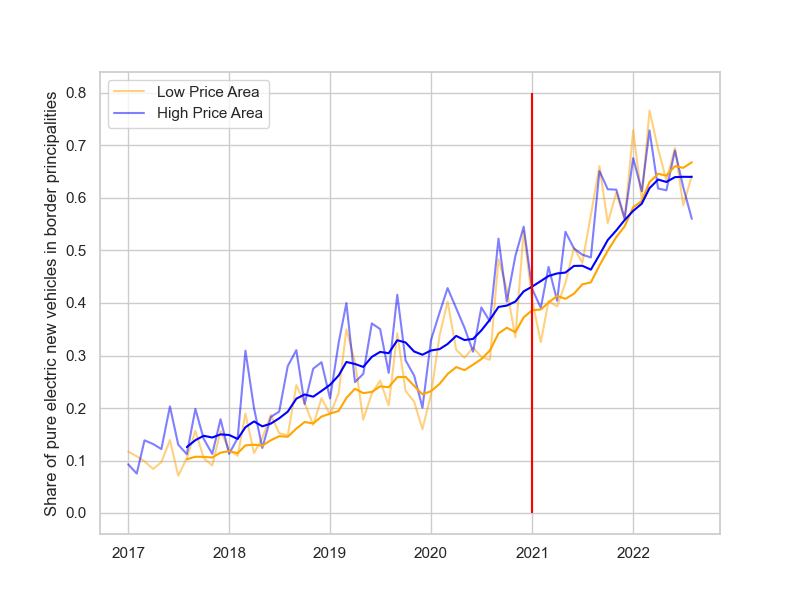}
      \caption{Limited to municipalities along the border of the high price areas (NO1, NO5) and low- price areas (NO3). }
      \label{fig:shareElectricBorder}
     \end{subfigure}
     \caption{The monthly average share of new registered electric vehicles in the area that experienced high electricity prices after 2021 (NO1, NO2, NO5) and the share in the area that continued to have low electricity prices (NO3, NO4). A smoothed moving average line is superimposed on the monthly data in order to show trends.}
    \end{figure}

\begin{figure}
  \begin{minipage}{.48\textwidth}
    \centering
    \includegraphics[width=1\linewidth]{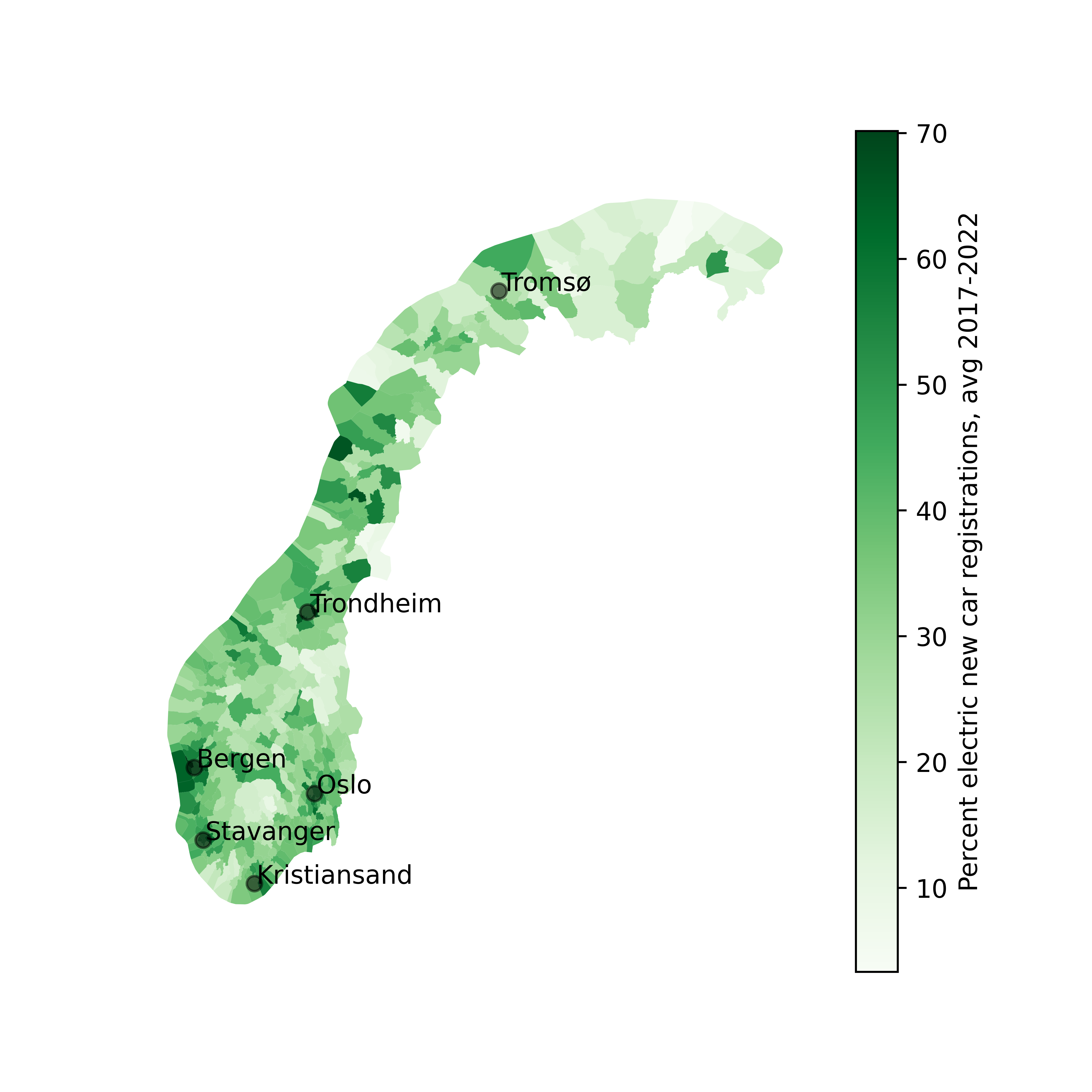}
    \caption{The figure shows the average proportion of fully electric car registrations by county for the years 2017 through september of 2022. Cities and areas with higher population density tend to have higher electric car penetration.}
     \label{fig:mapPercElec}
   \end{minipage}
   \begin{minipage}{.48\textwidth}
    \centering
    \includegraphics[width=1\linewidth]{mapBorder.png}
    \caption{Counties in Norway along the border of the NO3 price area in the north and the NO1 and NO5 price areas in the south. In the analysis this sample of counties is used to help identify the effect of electricity prices by improving upon the ignorability assumption of the county assignment.}
   \label{fig:mapBorder}
   \end{minipage}\qquad
  \bigskip
  \end{figure}

  A necessary assumption in order to obtain identification in a natural experiment study is that assignment into control and treatment groups should in practice be considered random--what is called the ignorability assumption. A difference-in-difference design loosens this assumption by differencing out time-invariant factors. However, an assumption of equal trends absent the treatment is necessary for difference-in-difference to provide unbiased causal estimates. Figure \ref{fig:shareElectric} divides the observations into two groups: The area that experienced high prices in 2021 and 2022, referred to as the high price area, and the area that saw low prices in 2021 and 2022, referred to as the low price area. The figure shows the average share of electric vehicles among new registered vehicles over time with a smoothed moving average term in order to show the trends in the series. The high price area, composed mostly of the southern Norwegian areas, have on average a higher penetration of electric vehicles. This could potentially be explained by a higher population density and more cities in these areas, including the Oslo Region. As figure \ref{fig:mapPercElec} shows, cities and densely populated areas tend to have a higher share of electric vehicles. However the trend over time before the divergence in prices appears roughly similar. After the divergence in prices at the beginning of 2021, the difference in penetration has decreased. 
  
  In order to try to strengthen the ignorability assumption between the two price areas, I also limit the analysis to the set of municipalities along the border between the low-price NO3 price area in the north, and the high-price municipalities in the NO1 and NO5 price zones. Figure ref{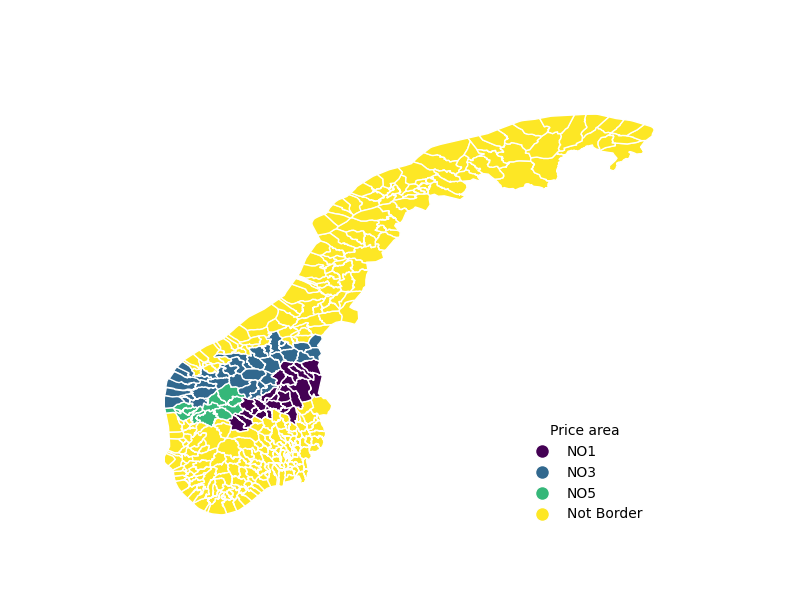} shows a map of the included municipalities in this subsample. These municipalities tend to be roughly similar, lacking any mid-sized or large cities or municipalities and having roughly similar climate and geography. Figure \ref{fig:shareElectricBorder} shows the monthly average share of new registered electric vehicles in the border region. The share of electric vehicle penetration in the high- and low-price areas are roughly similar. After prices diverged, the penetration appears to have accelerated in the low-price area.

\begin{figure}
        \begin{subfigure}{.48\textwidth}
          \centering
          \includegraphics[width=1\linewidth]{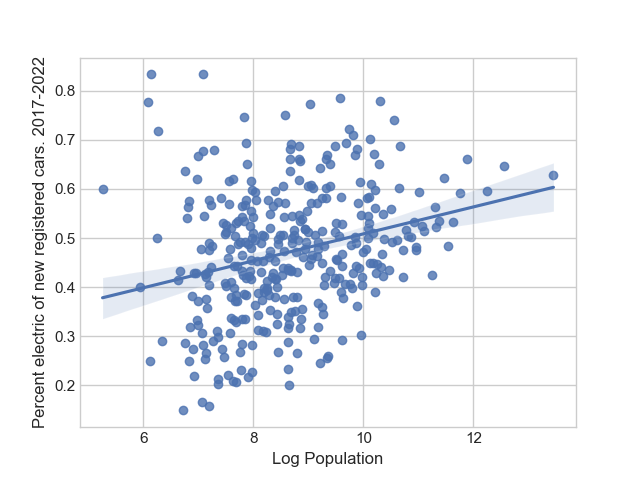}
          \caption{A scatterplot of the average share of electric cars registrations by county in the period 2017-2022 and the average income per capita by county.}
          \label{fig:logPercElec}
         \end{subfigure}
         \begin{subfigure}{.48\textwidth}
          \centering
          \includegraphics[width=1\linewidth]{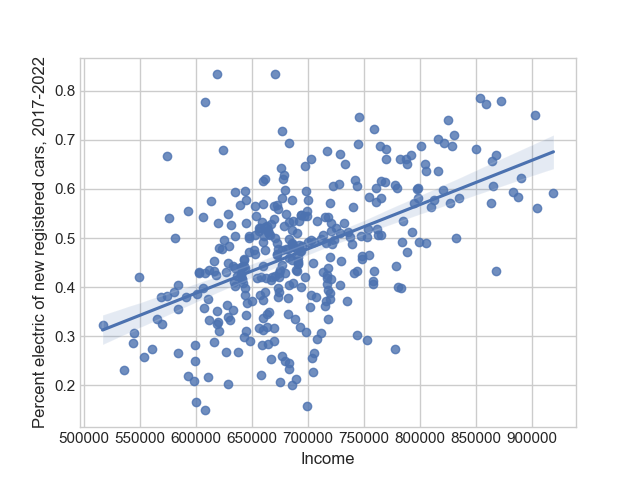}
          \caption{A scatterplot of the average share of electric cars registrations by county in the period 2017-2022 and the average income per capita by county.}
          \label{fig:incomeElectricScatter}
         \end{subfigure}
        \end{figure}

Several factors other than electricity prices may affect the electric car purchasing decision. Including such factors in the model can help correct for imbalance between the comparison groups and improving the precision of parameter estimates. Population density was already discussed above in relation to the location of cities in the different price areas. Population density could also play a role at a finer geographic level. Figure \ref{fig:logPercElec} shows a scatter plot of log average population of municipalities on the y-axis while the x-axis shows the average share of electric new vehicle registrations between 2017 through 2022. The variables are positively correlated, though with a large amount of variance.

Household income is another factor that might influence the decision to get an electric car. Figure \ref{fig:incomeElectricScatter} shows a scatterplot where the the median income of households in Norwegian kroners is shown on the x-axis and the average share of electric new registered vehicles is shown on the y-axis. Median income is positively correlated with the proportion of electric vehicles in a municipality. 

\section{Simple differencing estimates}

\begin{figure}
  \begin{minipage}{.48\textwidth}
    \centering
    \includegraphics[width=1\linewidth]{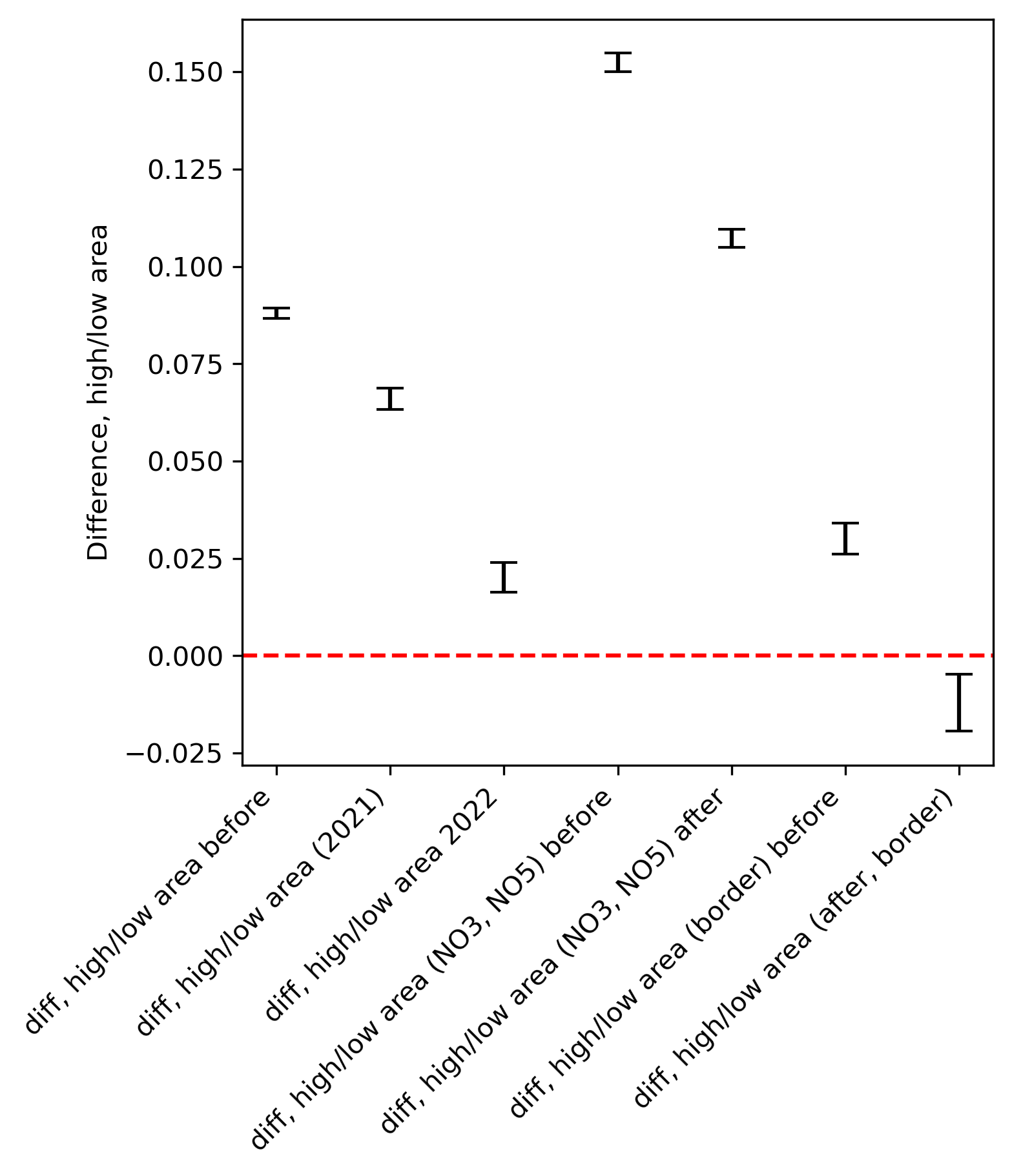}
    \caption{Simple differences in the probability of choosing an electric car between high and low price area, 2017-2022. The bands show approximate 68\% confidence intervals (+/- one standard error) for the estimates.}
     \label{fig:plotDiff}
   \end{minipage}
   \begin{minipage}{.48\textwidth}
    \centering
    \includegraphics[width=1\linewidth]{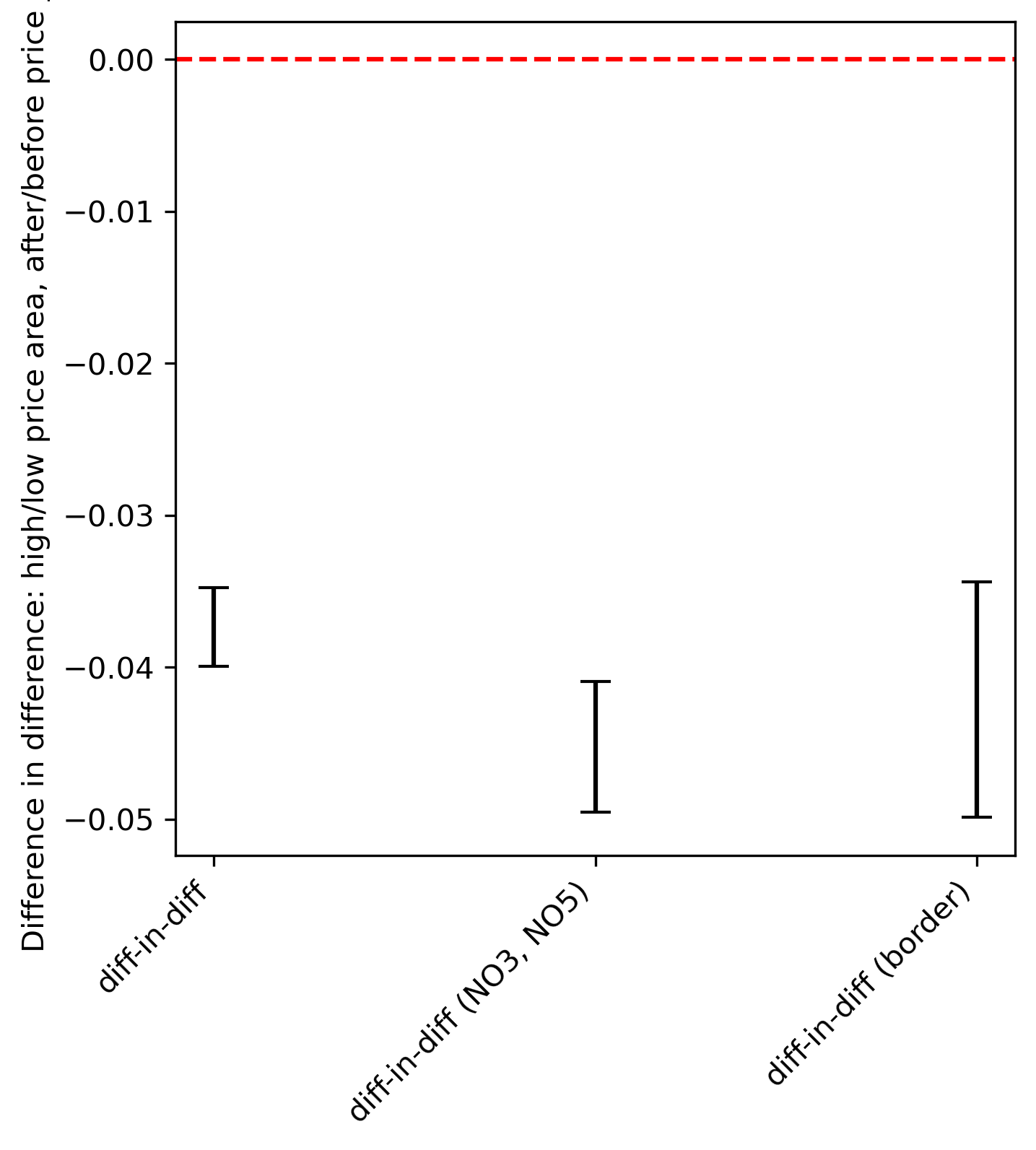}
    \caption{Simple difference-in-difference estimates between low- and high price areas and before and after prices diverged.}
    \label{fig:plotDiff_in_diff}
   \end{minipage}\qquad
  \bigskip
  \end{figure}

Figure \ref{fig:plotDiff} shows estimates of simple differences for the average proportion of electric new registered vehicles between the high and low price area (high price area - low price area.) The intervals represent approximate 68\% confidence intervals of the estimates (+/- 1 standard error). The figure shows six estimates. The first is a difference for the entire data between the high price areas and low price areas before 2021. Here we get a coefficient of approximately .075, which we can interpret as meaning that the proportion of electric cars sold in the high price area is 7,5\% higher than in the low price area on average. As discussed, this mainly reflects structural factors between the southern and northern price areas like population density and income. The second and third intervals from the left represents the difference in 2021 and 2022, when prices diverged. We see that the difference between the south (high price) and north (low price) declined. 

In the fourth interval from the left, the difference is limited to the mid-Norway NO3 area (low price) and the western NO5 area (high price) before 2021. These two areas are adjacent and they exclude the Oslo metropolitan area in NO1. Each contains a medium sized regional city (Bergen on the west coast and Trondheim in mid-Norway.) Here the estimate is approximately .15. The fifth interval from the left shows the difference between the NO5 area and the NO3 area in 2021 and 2022. Again the magnitude of the difference has been reduced.

The sixth estimate from the left looks at the difference only between high and low municipalities that are in the border region of the price areas (as shown in figure \ref{fig:mapBorder}). The last interval from the left shows again results from the border subsample, this time only for the years 2021 and 2022. The interval has gone from being centered around .025 to approximately -.015.

Figure \ref{fig:plotDiff_in_diff} shows results from simple difference-in-difference estimates. The differences are taken between the high and low areas before and after the start of 2021. The first interval is calculated from data from the entire Norway and is centered at approximately -.037. Limiting the comparison to the NO3 and NO5 price areas, the estimate is somewhat larger in magnitude as shown in the second-from-the-left interval. The third interval represents the estimation for the border region, with a magnitude of approximately -.035 to -.05. All of these estimates provide support for a significant effect of higher electricity prices on the probability of electric car purchase roughly in the magnitude of between 3\%-5\%.  

  The results of the simple differencing are suggestive of a significant effect of higher prices on electric car adoption. In the following sections, results are presented from difference-in-difference regression models that reduce bias and improve the precision of estimates by explicitly modelling a time trend and controlling for variables such as population and household income. 

  \subsection{Regression discontinuity regressions}

  \subsection{Simple difference-in-difference regressions}

  In this section I present results from difference-in-difference regressions that explicitly models a time trend and estimates the effects of the diverging prices under asssumptions of both a constant parallel trends as well as a trend with varying slopes. A starting point can be a model as written below: 
  
  \begin{equation}
    \label{eq:diff_in_diff}
        I^E_{ict} = \alpha + \tau timeDays_{i} + \gamma high_c + \zeta after_t + \eta high_c*after_t  + \mathbf{X \beta_{c}} +\lambda group_{i} + \epsilon_{ict}
  \end{equation}
  
Here the left-hand side variable is a binary variable, $I^E_{ct}$, which is equal to 1 if a new registered vehicle is electric and 0 otherwise. The indicator variable is indexed both by time, in days, and by the municipality, c. $timeDays_{t}$ is a calendar time variable measured in days since the first day of 2017 and the coefficient $\tau$ represents the estimated slope of the linear time trend. $high_c$ is an indicator variable indicating whether a certain municipality is located in the price zone that experienced high prices after the start of 2021 and the coefficient $\gamma$ can be interpreted as the group difference between high and low price areas in the area before prices diverged. Similarily $after_t$ is an indicator variable that is equal to 1 if a new vehicle registration happened after prices diverged and the coefficient $\zeta$ represents the average difference in the proportion of electric new vehicle registrations before and after prices diverged in the low-price area. 

The coefficient $\eta$ on the interaction term $high_c*after_t$ then represents the difference-in-difference estimator. This is the change in the relative difference between the high and low areas, before and after prices diverged. Finally, a vector of municipal level controlling variables are represented by the term $\mathbf{X \beta_{c}}$. In particular, measures of population ($population\_st$) and after-tax income ($income\_at_{st}$). The $st$ indicates that both variables have been standardized by subtracting the mean value and dividing by a standard deviation. The inclusion of these variables will not necessarily affect the point estimation of the difference-in-difference estimator since these variables are time-invariant. However, by controlling for factors that likely lead to imbalance between the comparison groups, the inclusion of these variables can improve the precision of the difference-in-difference estimator. This simple form of a difference-in-difference regression assumes that the slope of the time trend is constant both between high price areas and before and after prices diverged. 

The first three columns of table \ref{tbl:ddregressions} shows results from the model. The first column shows results from all of Norway. The second column restricts the sample to the price areas NO3 and NO5 and column three limits the sample to municipalities in the border region to the high and low price zones as shown in figure \ref{fig:mapBorder}. In the sample that includes all of Norway, the difference-in-difference coefficient has a point estimate of approximately -.025, with an approximate 95\% confidence interval of between -.03 to -.021. In the subsample that includes only price areas NO3 (low price) and NO5 (high price), the magnitude of the effect increases to approximately -.034. In the subsample of municipalities around the border, the point estimate is estimated to be -.046, with a 95\% confidence interval of between -.032 and -.060. These estimates support the hypothesis that higher electricity prices reduced the proportion of electric car purchases by a magnitude of between 2 and 5\%.


  \begin{table}
    \caption{}
    \label{tbl:ddregressions}
    \begin{center}
    \begin{tabular}{llllll}
    \hline
   & \multicolumn{3}{c}{Fixed trend} & \multicolumn{2}{c}{Varying trend} \\
                                          & All Norway & NO3 and NO5 & Border regions & All Norway & Border regions  \\
    \hline
    afterPriceJump                        & 0.0376                  & 0.0394                   & 0.0668                      & -0.2991                   & -0.4911                        \\
    & (0.0025)                & (0.0036)                 & (0.0066)                    & (0.0188)                  & (0.0424)                       \\
highPriceArea                         & 0.0413                  & 0.0565                   & 0.0522                      & 0.0413                    & 0.0522                         \\
    & (0.0013)                & (0.0028)                 & (0.0040)                    & (0.0013)                  & (0.0039)                       \\
timeDays                              & 0.0003                  & 0.0003                   & 0.0002                      & 0.0003                    & 0.0002                         \\
    & (0.0000)                & (0.0000)                 & (0.0000)                    & (0.0000)                  & (0.0000)                       \\
population\_st                        & 0.0591                  & 0.1164                   & 0.0683                      & 0.0592                    & 0.0721                         \\
    & (0.0005)                & (0.0026)                 & (0.0328)                    & (0.0005)                  & (0.0327)                       \\
med\_income\_at\_st                   & 0.0348                  & 0.0796                   & 0.0548                      & 0.0349                    & 0.0551                         \\
    & (0.0005)                & (0.0019)                 & (0.0027)                    & (0.0005)                  & (0.0027)                       \\
after:high          & -0.0257                 & -0.0342                  & -0.0461                     & 0.1655                    & 0.2355                         \\
    & (0.0024)                & (0.0040)                 & (0.0071)                    & (0.0209)                  & (0.0630)                       \\
timeDays:after               &                         &                          &                             & 0.0002                    & 0.0003                         \\
    &                         &                          &                             & (0.0000)                  & (0.0000)                       \\
timeDays:after:high &                         &                          &                             & -0.0001                   & -0.0002                        \\
    &                         &                          &                             & (0.0000)                  & (0.0000)                       \\
R-squared                             & 0.2113                  & 0.2381                   & 0.2371                      & 0.2118                    & 0.2396                         \\
R-squared Adj.                        & 0.2113                  & 0.2380                   & 0.2370                      & 0.2118                    & 0.2395                         \\
N                                     & 1022868                 & 228940                   & 64723                       & 1022868                   & 64723                          \\
AIC                                   & 1196342.17              & 264973.61                & 67717.68                    & 1195764.97                & 67508.06                       \\
R2                                    & 0.21                    & 0.24                     & 0.24                        & 0.21                      & 0.24                           \\
Adj. R2                               & 0.21                    & 0.24                     & 0.24                        & 0.21                      & 0.24                           \\
\hline
    \end{tabular}
    \end{center}
    \end{table}

\begin{figure}
    \begin{subfigure}{.33\textwidth}
      \centering
      \includegraphics[width=1\linewidth]{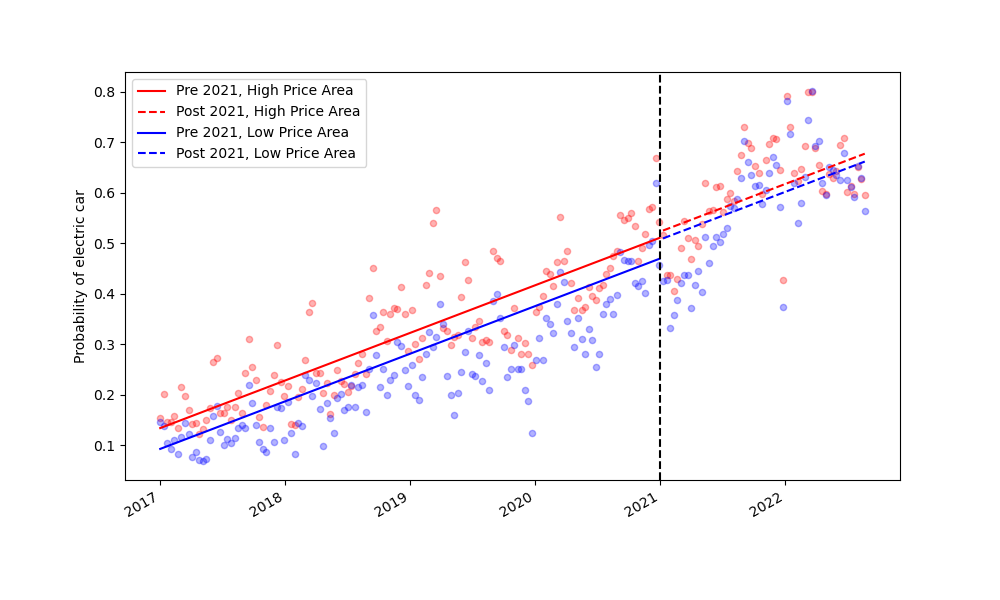}
      \caption{All of Norway.}
    \end{subfigure}%
    \begin{subfigure}{.33\textwidth}
      \centering
      \includegraphics[width=1\linewidth]{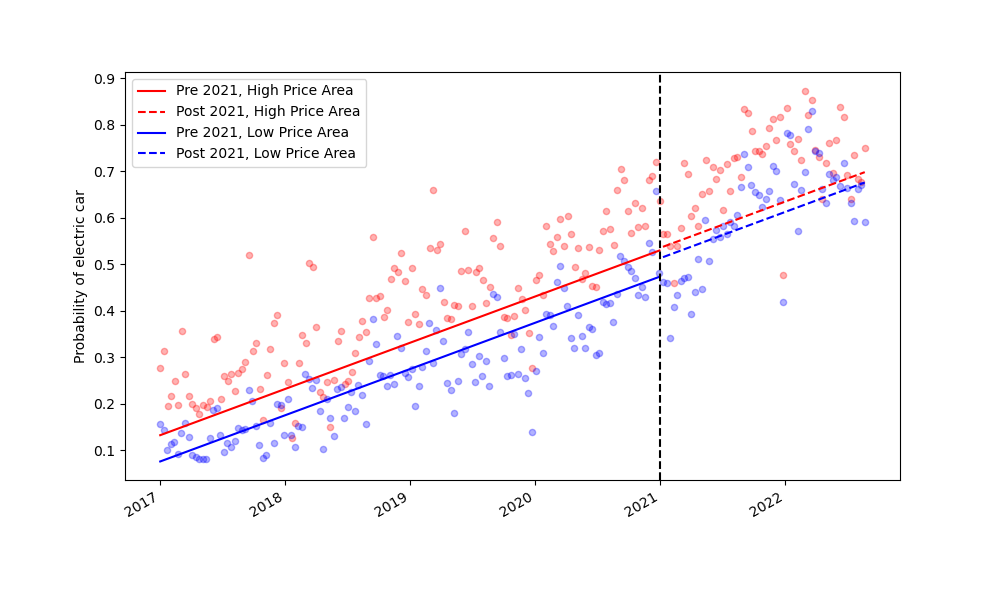}
      \caption{NO3 and NO5}
    \end{subfigure}
    \begin{subfigure}{.33\textwidth}
      \centering
      \includegraphics[width=1\linewidth]{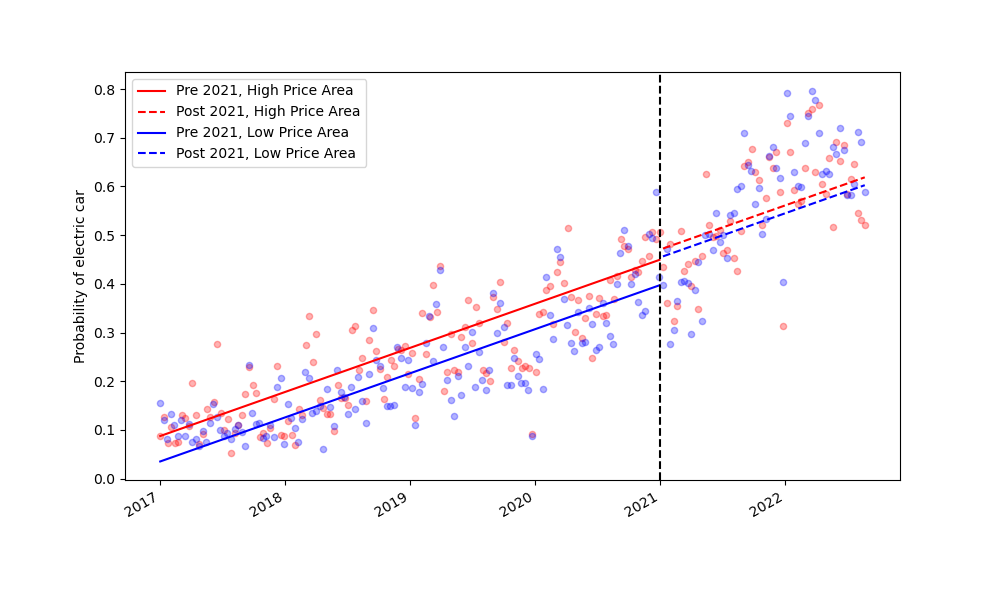}
      \caption{Border counties}
    \end{subfigure}
    \caption{Illustration of difference-in-difference regression estimates. }
    \label{fig:DDregressionPlots1}
    \end{figure}

The magnitudes of the results as well as the overall validity of the model are best evaluated visually. Figure \ref{fig:DDregressionPlots1} shows illustrations of the model results in the form of regression lines overlayed by the data as presented as average binned values. The red lines represent the estimated regression line for the high price area and the blue line represents the estimated regression line for the low-price area. The vertical dotted black line represents the start of 2021 when prices in the two areas diverged. In panel A results are shown when the sample included all new registrations in Norway, panel B shows results from  NO3 and NO5 price areas and panel C shows results from the subsample from the border region. The results are similar. The penetration of electric vehicles was initially larger in the high price area in southern Norway, while the level of the trend stayed steady in the high-price areas in the south, the level of the trend rose in the low-price area after prices diverged. 

Overlaying the results of the regression lines over the binned data suggests that the assumption of parallel trends does not hold, especially in the post-2020 data. I can loosen this assumption, though at the cost of loosing the clear interpretation of the difference-in-difference estimator.

\begin{figure}
    \begin{subfigure}{.48\textwidth}
      \centering
      \includegraphics[width=.8\linewidth]{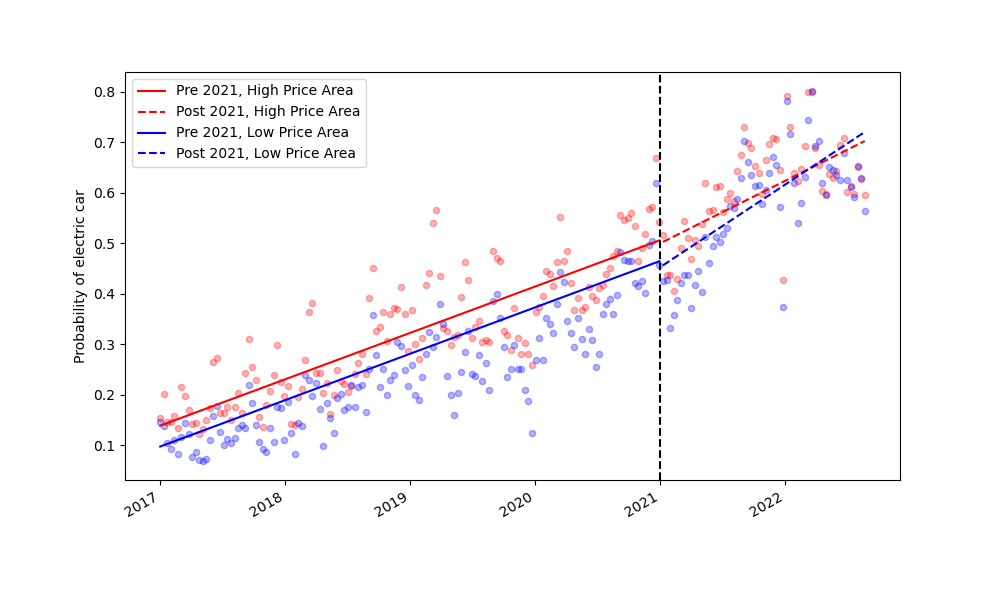}
      \caption{All of Norway.}    
    \end{subfigure}
\begin{subfigure}{.48\textwidth}
  \includegraphics[width=.80\linewidth]{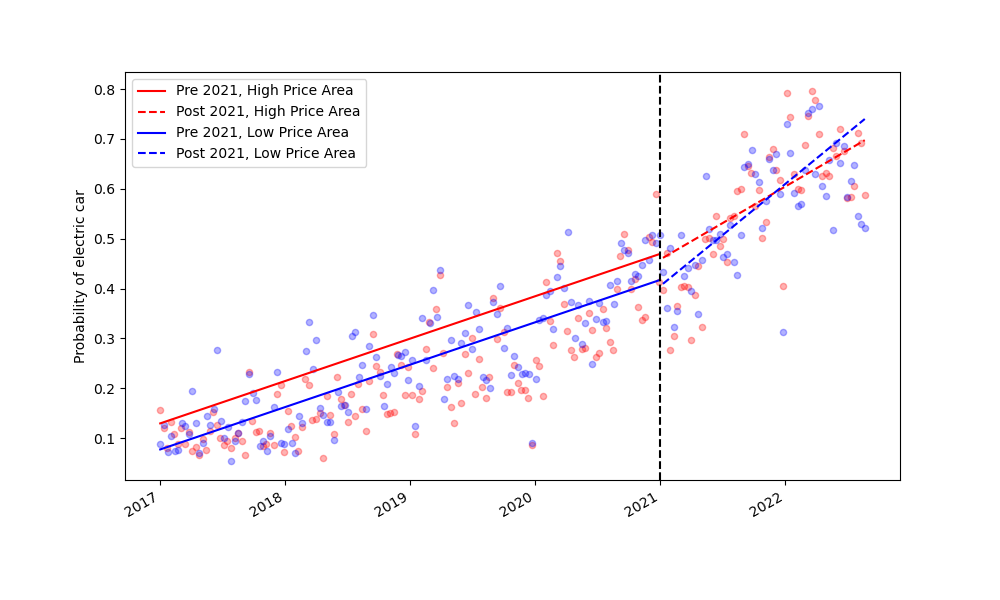}
  \caption{Border region.}
\end{subfigure}
\caption{Representation of model results allowing for post-2021 slopes to vary by high- and low price areas.}
\label{fig:DDregressionPlots2}
\end{figure}

The fourth column shows results where the assumption of constant slope of the time trend are loosened. In particular the interaction terms $timeDays*afterPriceJump$ and $timeDays*after*high$ are added to the regression. This allows the slope of the time trend to vary between high- and low-price areas after prices diverged in 2021. The coefficients in themselves are not easy to interpret directly, but the visualisation of the model results are presented in figure \ref{fig:DDregressionPlots2}. In panel A results for all of Norway is shown. In Panel B, results for the border region are shown. Now, instead of showing a discrete jump in the penetration of electric vehicles, the model shows a relative increase in the slope of the trend in the low-price area after prices diverged. This is likely a better representation of the dynamics of electric car choice as prices rose in the southern price areas but stayed low in the northern areas. Prices diverged gradually over time, thus estimating a single discrete jump is not realistic.

\begin{figure}
  \includegraphics[width=.8\linewidth]{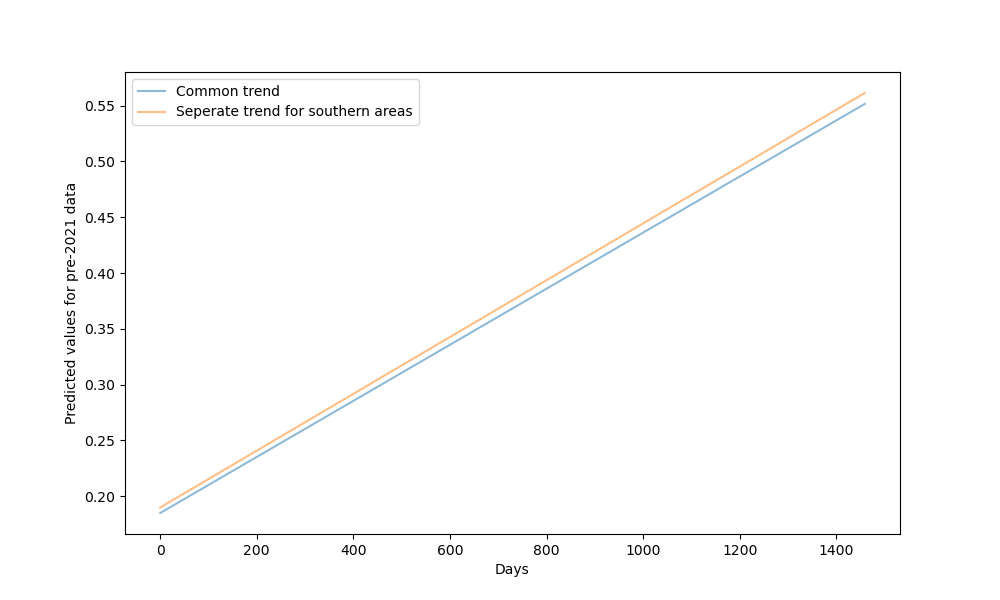}
  \caption{The figure shows predicted values of a model fit on pre-2021 data with and without specifying varying trends between the southern and northern price areas holding all variables other than the days constant. Assuming a common slope in the pre-2021 period appears to be a reasonable assumption.}
  \label{fig:preTrendCheck}
\end{figure}

In this model I still make an assumption that the slope of the trends in the pre-2021 period for the southern and northern price areas are the same. I can test this assumption by running models on the pre-2021 data with both an assumption of a common time trend, and a model which allows for a varying time trend by including an interaction effect between the time variable and the indicator variable for the southern price areas (the "high price areas"). The results are best evaluated visually, since with over 1 million observations, even small, practically insignificant differences in trend are likely to still be statistically significant. Figure \ref{fig:preTrendCheck} shows predicted values from the two models for the 4 year period from 2017 through 2020, holding all variables except the time variable constant. The model where a seperate trend is estimated for the high price area has a slightly steeper slope, but the assumption of a common slope appears reasonable. 

\section{Counterfactual estimate of magnitude}

In this section, I make use of simulations from the above model results to make some approximate calculations about the magnitude of the effect of high electricity prices on electric car purchasing. I use the model that allows for varying slopes of the time trend after prices diverge. I choose this model both because it appears to fit the data better than the model with fixed slopes of the trend. This coefficients in this model also do not lend themselves to a direct and clear interpretation of the magntitudes of the effect.

\begin{figure}
  \begin{minipage}{.48\textwidth}
    \centering
    \includegraphics[width=.8\linewidth]{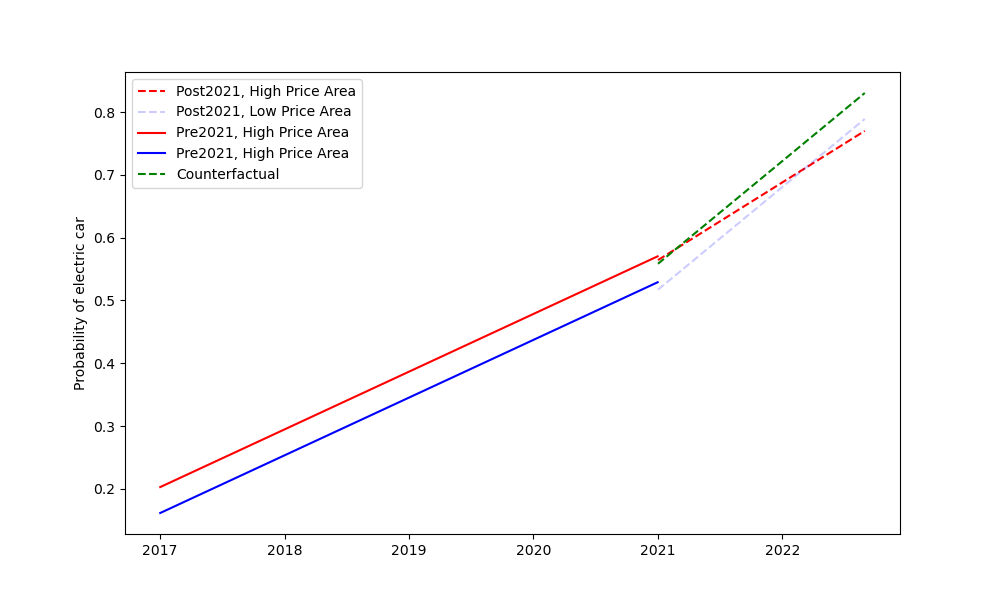}
    \caption{Illustration of counterfactual. The assumption is that the change in slope seen in the low-price area would also be experienced in the high-price area if prices were low.}
    \label{fig:CFPlot1}
  \end{minipage}
  \begin{minipage}{.48\textwidth}
    \includegraphics[width=.8\linewidth]{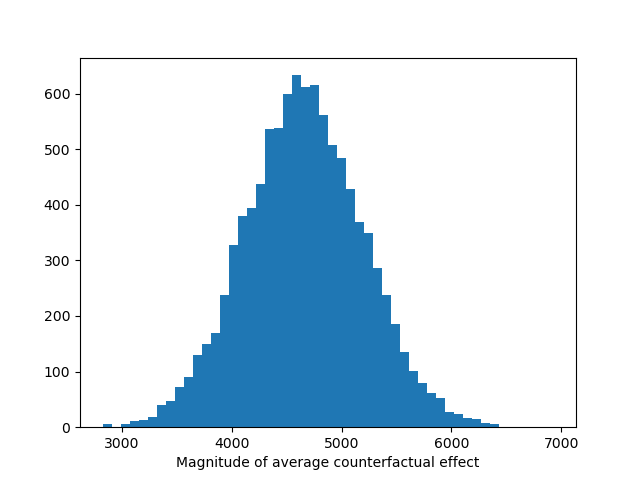}
    \caption{The estimated distribution of the average counterfactual effect of higher prices. The distribution is generated by sampling from a multivariate normal distribution based on the point estimates and variance-covariance matrix of the model estimation.}
    \label{fig:CFEffect_dist}
  \end{minipage}
  \end{figure}

The magnitude I wish to calculate is the difference between the number of electric cars sold in the high-price area after prices diverged and an estimated counterfactual of the number of electric cars sold if prices had been as low as in the northern price areas. Figure \ref{fig:CFPlot1} provides a visualisation of that counterfactual. In the figure, the red dotted line represents the average probability of choosing an electric car in the southern (high price) areas. The green line represents the average counterfactual estimate for the probability of choosing an electric car had prices been as low as in the northern area. The main assumption is that the level at the beginning of 2021 is adjusted for the southern price area while the slope is assumed to be equal to that of the northern (low-price) area. 

The counterfactual is represented by the equation in \ref{eqn:counterfactual1}. Here the probability of a certain registered new vehicle, i, being an electric vehicle is represented by $p_i^C$, with the superscript \textit{C} representing that this is the counterfactual. $I^H$ represents the indicator variable for the high price area, which here will always be equal to one since the counterfactual only includes observations in the southern, high-price area, this will always be equal to 1. $\gamma^H$ represents the estimated intercept adjustment for the high price area. $\hat{\delta^L}$ represents the slope parameter on the time trend for the northern, low-price area. 

\begin{equation}
  p_i^C = \hat{\alpha} + \hat{\gamma^H I^H} + \hat{\delta^L} t_i + \mathbf{X_i \hat{\zeta}}
  \label{eqn:counterfactual1}
\end{equation}

The magnitude of the effect, $M^C$ is then calculated as the sum of the counterfactuals for each observation, i, in the high price area after the prices diverged minus the actual number of electric cars sold in that period. This is shown in equation \ref{eqn:counterfactual2}

\begin{equation}
  M^C = \sum_i p_i^C - \sum_i y_i^H
\label{eqn:counterfactual2}
\end{equation}

The model parameters are subject to uncertainty, and therefor the estimate of the magnitude should also be presented in relation to this uncertainty. This is done by simulating from a multivariate normal distribution with the mean parameters and covariance matrix taken from the model results, as shown in equation \ref{eqn:counterfactual3}, where $\mathbf{\beta^S}$ is a vector of $k$ random variables corresponding to the parameters in the model. $\mathbf{\hat{\beta}}$ represents a vector of the the point estimates of parameters from the regression model that represent the mean values. $\mathbf{\hat{\Sigma}}$ is the estimated variance-covariance matrix from the model. 

\begin{equation}
  \mathbf{\beta^S} \sim \mathcal{N_k}(\mathbf{\hat{\beta}}, \mathbf{\hat{\Sigma}}) 
\label{eqn:counterfactual3}
\end{equation}

The resulting distribution of the estimated counterfactual magnitude is shown in figure \ref{fig:CFEffect_dist}. The distribution is centered at just around 5000 vehicles. This can be interpreted to mean that, given the assumption that identification is correct, had the southern price area had as low prices as in the north, 5000 more vehicles would have been registered as electric rather than petrol, diesel or hybrid. In the period after prices diverged, a total of 242,000 vehicles were registered in the southern price area. Thus this represents an effect of roughly 2\%. This estimate is then in line with the estimates from the earlier difference-in-difference model where slopes where constrained to be constant.

These calculations are all based on in-sample predictions. At a technical level, it may be possible to create forecasts for the future and in turn create calculations of the effect on electric car penetrations. But these calculations would be subject to great uncertainty. First, this would involve making assumption about future electricity prices. A more fundamental problem is that predictions from this model will inherently be linear. This can be an acceptable simplification at penetrations of electric vehicle registrations that are at intermediate values. But assuming an S-shaped diffusion process, as penetration approach a saturation level--as they may soon do in Norway--linearity assumptions would likely be inappropriate. 

While the effects are noticeable, given the large difference in electricity prices in this period, the effects on electric car choice is nonetheless modest. Noteably, even though prices soared to rates that were at times 100 times the price from in the northern region, the trend in the southern Norwegian areas was still towards a higher proportion of electric vehicles, the rate of change was just less than it otherwise could have been.

\section{Conclusion}

How electricity prices affect the adoption of electric cars has been a little studied topic in the literature. This article presents evidence from a unique case where we can compare areas with large differences in electricity prices yet where institutional, political, cultural and climactic factors are largely similar. The evidence points to a measurable effect of higher electricity prices with a magnitude of 2-5\% less probability of purchasing an electric car in the high price area. 

Estimating a more direct relationship between price and probability of adoption--estimating a price elasticity for example--could be attempted at a purely technical level. But such an estimation would be subject to many uncertainties. Most notably, there are significant and unknown lags involved between when a decision is made to purchase a certain type of car, presumably involving taking into consideration electricity prices, and when that car is actually registered. Thus any estimation of the relationship between a specific price difference and the probability of adopting a car would involve some strong assumptions and in turn estimates with false precision. In this article I have instead chosen to estimate a magnitude between a high price area and a low price area.

Perhaps the most notable aspect of the results is not the result that higher electricity prices led to a lower probability of electric car adoption. This result would be predicted by the simplist of economic models and given the size of the price differences between comparison areas and the number of data points, it would have been surprising if no effect were found. Rather, the more significant result is how modest the effect was. In the period with record high electricity prices in the south, penetration of electric vehicle purchases in the new vehicle market continued to increase, only at a somewhat slower rate than in the northern price area with low prices. 

A reasonable explanation for this modest effect is that consumers, perhaps rationally, saw the high prices in 2021 and 2022 as temporary. In making a buying a decision, a rational consumer for a vehicle that they may own for a decade or more, will discount even large, but temporary increases in the price of electricity. However, this would seem to represent a change in behavior relative to the evidence for how consumer expectations are formed for future gasoline prices. \citet{anderson_what_2013} find in an analysis of survey data that consumers on average have beliefs consistent with a no-change forecast. Thus, on average, a consumer faced with higher gasoline prices will tend to forecast those high prices into the future when making consumption decisions. 

Another explanation is that a combination of factors such as an inherent efficiency of an electric motor, state subsidies for electric cars, and an increasing availability of a variety of affordable electric vehicles dominates the effects of electric prices. No matter the underlying mechanism, the implication of the results seem to be that higher--even much higher--electricity prices are unlikely to substantially impede the electrification of passenger vehicles. 

\section{Software and replication}

All computation and analysis in this article is done using the Python programming language \citep{van_rossum_python_2009} and the Numpy library \citep{2020NumPy-Array} for numerical computation. The library Pandas \citet{mckinney_data_2010} is used for data management, transformation and table creation. The libraries Matplotlib \citep{hunter_matplotlib_2007} and Seaborn \citep{waskom_mwaskomseaborn_2017} are used for visualization. The libraries Statsmodels \citep{seabold_statsmodels_2010} and Scikit.learn \citep{pedregosa2011scikit} are used for regression modelling. Geopandas \citep{jordahl2014geopandas} is used to work with geographic data and map creation.

Data and code are available upon request.

  \bibliographystyle{chicago}
  \bibliography{cars}
  
  

\end{document}